# Nuclear Fusion: Bringing a star down to Earth

A. Kirk

CCFE, Culham Science Centre, Abingdon, Oxon, OX14 3DB, UK

## Abstract

Nuclear fusion offers the potential for being a near limitless energy source by fusing together deuterium and tritium nuclei to form helium inside a plasma burning at 100 million kelvin. However, scientific and engineering challenges remain. This paper describes how such a plasma can be confined on Earth and discusses the similarities and differences with fusion in stars. It focusses on the magnetic confinement technique and, in particular, the method used in a tokamak. The confinement achieved in the equilibrium state is reviewed and it is shown how the confinement can be too good, leading to explosive instabilities at the plasma edge called Edge Localised modes (ELMs). It is shown how the impact of ELMs can be minimised by the application of magnetic perturbations and discusses the physics behind the penetration of these perturbations into what is ideally a perfect conducting plasma.

# 1. Introduction

Nuclear fusion is the process that powers the Sun and all other stars. The ability to harness this technology has many attractions for future electrical power plants including almost unlimited supplies of the fuel, more than 10 000 years on Earth, high energy density base load electrical generation, $CO_2$ free production and no long lived radioactive waste.

In principle nuclear fusion is easy; it works by fusing two light nuclei to form new nuclei, which have less mass than the initial two; the missing mass is released as energy. The problem comes because in order to make the nuclei fuse we need to bring them close enough together – they need sufficient energy to overcome the Coulomb barrier. One way to do this is to heat the particles but, depending on the nuclei, the temperatures required are in the range of 10-100 million K.

In the Sun the fusion processes is confined by gravity. The Sun uses the so called proton cycle;

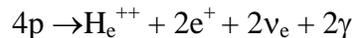

Through a series of steps 4 protons fuse to produce a Helium nucleus – the processes is governed by the weak nuclear force and hence is slow and not very efficient. On average one proton fuses every billion years and a 1m$^3$ volume of the Sun only produces 30 W of heat, which is less than the average human. The Sun works because it is so big. The advantage of such a slow process is that the lifetime of the Sun is sufficiently long (~ 10 Billion years) that humans have been able to evolve sufficiently to criticise its efficiency! For a power plant on Earth a faster process is required; one using the strong nuclear force. Hence the fuels typically used are heavy isotopes of hydrogen namely; Deuterium (composed of one proton and neutron) and Tritium (composed of one proton and two neutrons.) The process is;

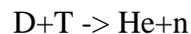

With the Helium nucleus (or α particle) carrying away 3.5 MeV of kinetic energy and the neutron 14.1 MeV of energy.

The peak fusion rate coefficient is ~$10^{25}$ times greater (~$10^{-21}$ m$^3$s$^{-1}$) for D-T compared to p-p (~$10^{-46}$ m$^3$s$^{-1}$), with a peak at temperatures of ~ 100 million K. To give a comparison with the Sun – if the Sun used the DT fusion cycle it would have lasted for <

1 second – a quite amazing second but we wouldn't have been around to appreciate it! Deuterium occurs naturally in water; while tritium needs to be bred from lithium using the neutron released from the DT fusion reaction in the process n+Li->He+T.

The fuel needed to supply the life-time electricity needs of an average person in an industrialised country can be provided by the D from half a tub full of water (250 l) and the T (15g) produced from the Li contained in a laptop battery (30g).

In this paper we will look at how to confine a fusion plasma, and in particular we will concentrate on the magnetic confinement technique and in particular the "Tokamak". The Tokamak is a device that was invented in the 1960s in the Soviet Union and uses magnetic fields to confine a plasma in the shape of a torus. We will investigate the equilibrium state of this tokamak plasma, how the high confinement achieved leads to explosive instabilities at the edge of the plasma and how these events can be tamed. To start with we will investigate an analogous example closer to home.

## 1.1 A Domestic analogue

Imagine putting a saucepan of water onto a hob and turning on the heat – eventually the water will start boiling, turbulence at the surface leads to steam carrying away some of the input energy – this is equivalent to what we will describe as a Low confinement mode (L-mode) of a tokamak plasma, where the edge turbulence carries away particles and heat, which reduces the overall confinement of the plasma.

To increase the heating efficiency when cooking we would put a lid on the pan – the losses from the pan decrease and the heating can be turned down while still maintaining the water temperature at boiling point – the cooking efficiency has been increased – the tokamak plasma has an equivalent to this – it is call the High confinement mode (H-mode), in which an insulating barrier forms at the edge of the plasma and the confinement time of the energy in the plasma effectively doubles.

The analogy continues because with a covered saucepan unless the heating is accurately adjusted the pressure of steam under the lid builds up until it is greater than the weight of the lid. Depending on the rate of heating, either the pan boils over, which is a plasma disruption or if the heating is not too great, the pressure increases just enough to raise the lid, which lets a bit of steam out and then reseals – this pressure relaxation cycle of pop–pop-pop from the saucepan lid is equivalent to what we will describe as an explosive instability in tokamaks called an Edge Localised Mode (ELM).

The analogy has one final part: In the kitchen to avoid this constant clattering of the lid– depending on the sophistication of your pan – you can either adjust a steam vent in the lid – or do like I do and stick a spoon under the lid – to allow just a little steam out to stop the annoyance but still maintain the good cooking efficiency. As we will discuss, a similar technique is used in tokamak plasma to minimise the effect of ELMs or remove them altogether.

## 2. Confining a Fusion plasma

In order to make fusion a reality you need to confine enough particles with sufficient energy for a long enough period – this is encapsulated in the so called Lawson criteria [1], given by the triple product

$$n.T.\tau_E > 3.10^{21} keVsm^{-3}$$

where $n$ is the density of the fuel, $T$ the temperature in keV and $\tau_E$ is the energy confinement time. $\tau_E$ is a parameter used to describe the rate of energy loss from the system and is equal to the energy density divided by the power loss density. There are several ways to achieve the Lawson criteria but all have to be at sufficient temperature for fusion to occur i.e. $10 < T < 100$ keV. At these temperatures the fuel is no longer in a solid or gaseous states it becomes a plasma. So the question is how to confine this hot plasma? Stars use gravity – but this is not an option on Earth. If the particles are compressed fast enough they can be contained for a short period of time by their own inertia. Inertial confinement fusion attempts to initiate nuclear fusion reactions by heating and compressing a fuel target, typically in the form of a pellet that most often contains a mixture of deuterium and tritium [2][3] and hence achieves the Lawson criteria using a high density (~$10^{26}$ m$^{-3}$ ) to compensate  for the low confinement time (~200 ps) -  such methods include laser induced fusion (see [4] and [5] and references therein).

The focus of this paper from now on will concentrate on another method, which uses the fact that charged particles can be confined using a magnetic field. This technique uses more moderate densities (~ $10^{20}$ m$^{-3}$) and longer confinement times (~1 - 10 s).

## 2.1 The principles of magnetic confinement

Charged particles gyrate around a magnetic field line, with fast motion parallel to the field and slow diffusion perpendicular to the field, typically due to inhomogeneities in the field or collisions. The simplest system would be a solenoid field containing a cylindrical plasma (see Figure 1a). Whilst the particles would be well confined in the cylinder there would be substantial end losses. Attempts to reduce these losses have been explored by changing the field shape at the end of the solenoid to produce a "magnetic bottle" [6] but to date the confinement times required have not been achieved. An alternative way of avoiding the end losses is to bend the cylinder around to form a torus (see Figure 1b).

In such a configuration the solenoid field coils become toroidal field coils and they create a so called toroidal magnetic field ($B_T$), which is in the toroidal direction denoted as $\phi$.

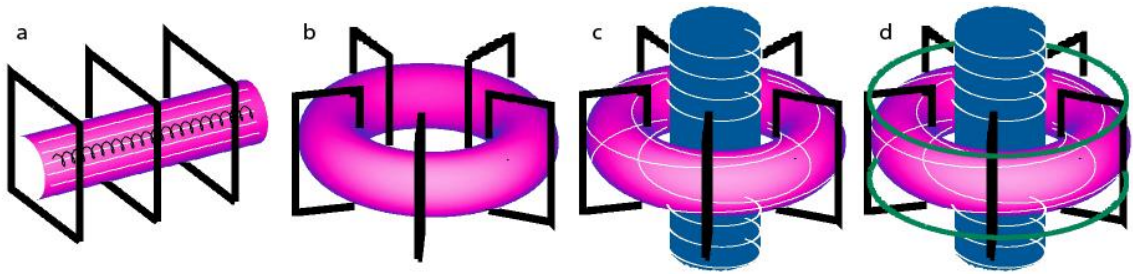

**Figure 1** Magnetic confinement of charged particles in a plasma a) in a cylinder, b) in a torus c) with an additional induced toroidal current in the plasma and d) with additional coils providing a vertical field.

The first problem that arises is that, because the coils are closer together on the inside of the torus than on the outside, the magnetic field is now stronger on the inside than the outside and the magnetic field varies as $B_T \propto 1/r$ , where $r$ is the radius of the torus. To understand the implications of this, firstly consider a positive and negative particle at the same location with the same speed in a constant magnetic field. At the outset the positive particle is launched upwards and the negative one downwards perpendicular to the magnetic field. Both particles move in a circle that coincides with one another. While there is an instantaneous electric field generated when the particles are at opposite sides of the circle the time averaged electric field is zero. Now consider the same two particles in a non-uniform field, similar to the one encountered in the torus and assume that the particles start at a large value of the plasma radius ($r$). As the particle is deflected in the field and moves towards smaller *r*, the magnetic field increases and the radius of curvature of the circle it is describing decreases. This leads to a net movement of the particles in the vertical direction with the positive particles moving upwards and the negative particle moving downwards (see Figure 2). This produces a time averaged net electric field in the vertical direction. The effect of this electric field is to create a so called $\vec{E} \times \vec{B}$ drift of the particles; the particle acquire a drift velocity ($v_d$) given by

$$v_d = \frac{\vec{E} \times \vec{B}}{B^2}$$

that is directed radially outwards. This leads to a loss of particles radially out of the plasma and the confinement time would be insufficient to achieve the Lawson criteria.

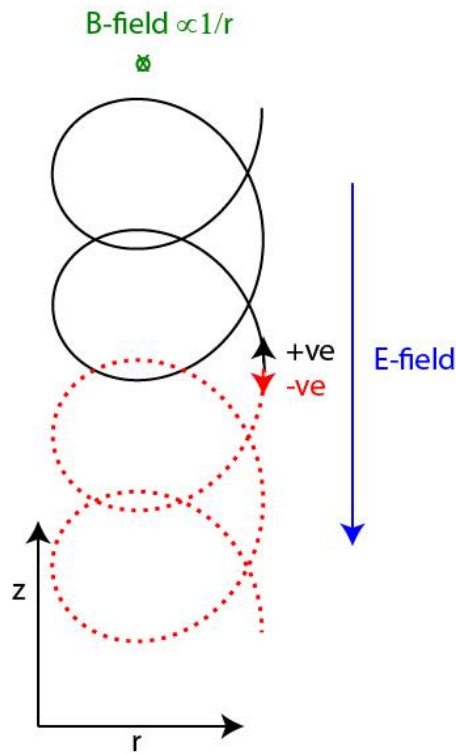

**Figure 2** Production of a vertical electric field due to the motion of charged particles in a radially varying magnetic field.

To remove this electric field, a helical transformation of the particles is required as they travel around the torus in the toroidal direction. Such a transform can be achieved in two ways. The first is to arrange the toroidal field coils in a helical shape, which is the technique employed in Stellerators [7]. The second way is to induce a current in the plasma in the toroidal direction. This then creates a magnetic field, which is perpendicular to the current. The total magnetic field is then helical i.e. it corkscrews around the plasma (Figure 1c). This is the technique employed in tokamaks [8][9], in which the plasma is contained in a doughnut-shaped vacuum vessel by externally applied magnetic fields and by an electrical current driven through the plasma. The tokamak was invented in the Soviet Union during the 1960s and soon adopted by researchers around the world. It is the most developed magnetic confinement system and will be the technique described in the remainder of this paper.

## 2.2 Magnetic confinement in tokamaks

The current that is required in a tokamak plasma is initially induced by transformer action. The primary coil of the transformer, called the solenoid, is inserted in the centre of the torus (see Figure 1c), with the plasma acting as the secondary coil of the transformer. The current in the solenoid (or primary) is ramped and induces a current of up to several MA in the plasma.

The plasma current ($I_P$) produces a so-called "Poloidal" magnetic field ($B_P$), which is in a plane that is at right angles to the toroidal field. The combination of the toroidal field from the external coils and the poloidal field from the plasma current produces a helical magnetic field. Every time the magnetic field goes once around toroidally, it is shifted by an angle ι by the poloidal field (Figure 1c). This angle is called the rotational transform.

The greater the plasma current, the larger the rotational transform ι. The number of times a field-line goes around toroidally for one poloidal turn is called the safety factor, q, and is given by;

$$q = 2\pi/\iota$$

If $q$ is a rational number $m/n$, then the field-line will join onto itself after $m$ toroidal turns and $n$ poloidal turns. However, if $q$ is irrational, then the field-line will trace out an entire surface. All field-lines lie on these surfaces, which are called flux surfaces (Figure 3a).

This is still not the end of fields required to produce a tokamak plasma that is in equilibrium; as the plasma is heated the pressure inside the plasma increases and the plasma expands radially outwards. In order to stop this radial expansion, a restraining ($\vec{I}_p \times \vec{B}$) force has to be applied. A radially inwards force can be supplied by applying a vertical field using a set of coils applied above and below the mid-plane in a Helmholtz-like configuration (Figure 1d).

In this magnetic configuration the extent of the plasma is limited by its interaction with the vacuum vessel structure that defines the tokamak, a so called "limiter plasma". Some of the early tokamaks used this configuration, including the Tokamak Fusion Test Reactor (TFTR), which was an experimental tokamak built at Princeton Plasma Physics Laboratory (in Princeton, New Jersey) around 1980 and produced 10.7 MW of nuclear fusion power in 1995 [10]. The disadvantage of the limiter configuration is that the interaction with the vessel wall occurs near to the core of the plasma. This leads to a cooling of the edge and core. In addition, impurity particles that are eroded from the wall due to the plasma interaction travel into the core, again cooling the core as well as diluting the fusion fuel.

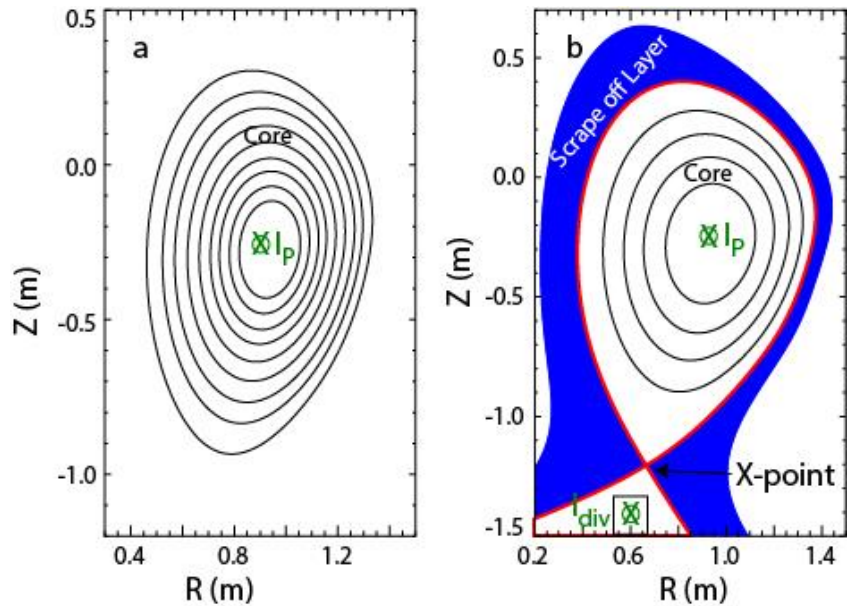

**Figure 3** Poloidal cross section (vertical cut at a given toroidal location) of a) a limiter plasma and b) a X-point diverted tokamak equilibrium showing the magnetic flux surfaces in the core of the plasma, the last closed flux surface (Red line) and the scrape off layer region.

Additional coils can be added to modify the flux surfaces at the edge of the plasma to divert the plasma that leaves the confined region towards the so called "divertor" targets. The function of the divertor is to extract the heat and particles released from the plasma, in effect acting as an exhaust system. It is made of materials, typically graphite or tungsten capable of coping with steady state power fluxes up to 10 $MWm^{-2}$.

Introducing a coil at the bottom of the vessel (see Figure 3b) that carries a current ($I_{div}$) in the same direction as the plasma current ($I_P$) produces a point at which the poloidal magnetic field is zero while the toroidal field remains unchanged. Field lines at this so called X-point, travel purely toroidally around the tokamak. This then divides the plasma into a core (closed magnetic flux surfaces) and a scrape off layer (open flux surfaces) region. The last closed flux surface (LCFS) defines the edge of the confined region of the plasma, particles that cross this surface follow the flux surfaces to the divertor. The divertor is a part of the vacuum vessel that is armoured in order to be able to cope with the heat fluxes leaving the plasma. The fact that the interaction of the plasma with material surfaces is now more remote leads to less impurities in the plasma and allows higher temperatures to be sustained at the plasma boundary. In 1997, using such a configuration, the Joint European Torus (JET), which is located at the Culham Centre for Fusion Energy, performed experiments in DT and set a record of 16.1 MW for the amount of fusion power produced [11].

## 2.3 Tokamak plasma equilibria

This set of coils can now produce a magnetic configuration of the plasma that is said to be in equilibrium: Charged particle gyrate around the field lines, but are free to move along them. In principle it is possible to derive all the plasma phenomena from the behaviour of individual particles and their interactions, however, the number of particles is so high that in practice such studies are limited. Instead the plasma is usually described as an electromagnetic fluid, which is encapsulated in MagnetoHydroDynamics (MHD). For a description of the application of MHD to tokamak plasmas see [12] and references therein. Such a fluid description allows the macroscopic behaviour of the plasma to be simulated without having to know the position and velocity of the individual particles. The so-called ideal MHD plasma description treats the plasma as being perfectly conducting. The resistivity of a plasma typically scales as $T^{-3/2}$ and hence the resistivity is low in the core of a fusion plasma and here the plasma can be treated as being a perfect conductor. These assumptions of ideal MHD dictate the connection between magnetic field lines and the plasma, in a sense tying the fluid to the magnetic field lines. As we will see later these assumptions break down near to the edge of the plasma where the temperature is lower and hence the resistivity higher.

For a plasma in equilibrium (i.e. no net force on the plasma) the equations of ideal MHD dictate that

$$\vec{j} \times \vec{B} = \nabla p$$

i.e. pressure gradients within the plasma are sustained by currents flowing perpendicular to the field lines. This perpendicular current is called a diamagnetic current and is given by

$$\vec{j}_\perp = \frac{\vec{B} \times \nabla p}{B^2}$$

and is produced by the difference in the velocities between the electrons ($v^e$) and ions ($v^i$) i.e.

$$\vec{j}_\perp = en(v^i_\perp - v^e_\perp)$$

Where $n$ is the density of ions and electrons (assumed to be equal) and the ions and electrons flow in opposite directions. The fluid drift or bulk plasma rotation is in the ion direction ($v^i_\perp$). For the conditions present in a tokamak plasma $v^i_\perp < v^e_\perp$ due to the larger collisional retardation of the ion motion. Hence the pressure balancing current is produced mainly by the electron motion and this current flows in the opposite direction to the fluid drift. The electron velocity is given by

$$v^e_\perp = \frac{\nabla p \times \vec{B}}{enB^2}$$

and hence is strongly driven by the pressure gradients. In a perfect tokamak that has full toroidal symmetry, the so called axisymmetric configuration, the plasma equilibrium becomes a 2D problem that can be simplified to the Grad-Shafranov equation [13]. This equation is implemented into equilibrium solvers to calculate the flux surfaces, current density and pressure gradients within a plasma based on external magnetic field measurements (see for example [14][15][16]).

As we will discuss later the pressure gradients also lead to the generation of parallel currents, which have an important effect on the plasma dynamics, for now we will just discuss how these currents are produced. In a tokamak plasma the charged particles have a velocity component along the field line ($v_{//}$) and perpendicular ($v_\perp$) to it. The magnetic moment of a particle is defined as

$$\mu = \frac{m v_\perp^2}{2B}$$

which is a conserved quantity. Since a magnetic field does no work on a particle the kinetic energy of a particle is constant hence $v^2 = v_{//}^2 + v_\perp^2$. This can be re-written as

$$v_{//}^2 = v^2 - \frac{2B\mu}{m}$$

In a tokamak $B \propto 1/r$, therefore, for a particle that starts at large r moving towards smaller $r$, as $B$ increases $v_{//}^2$ first decreases towards zero and, depending on the size of $B$ and $v$, can in fact become negative. Since $v_{//}$ does not become imaginary the particle must reflect. This configuration of magnetic field lines is called a magnetic mirror. This is created when the field lines converge, increasing the magnetic field line density, which means that approaching charged particles are retarded and can be reflected along their initial path. This mirror effect will only occur for particles whose velocity vector is within a certain range of angles to the field line and particles then become trapped in what is referred to as a banana orbit (Figure 4a) if

$$\left( \frac{v_{//}}{v_\perp} \right)^2 < \frac{B_{max} - B_{min}}{B_{min}}$$

Consider two such banana orbits which are touching (Figure 4b). If there is a density gradient, then there are more particles on the inner orbit than the outer one and so there is a net flow of trapped particles and hence a current is generated. The particles that are not trapped in these banana orbits are referred to as passing particles. In fact, both trapped and passing particle contribute to what is called the bootstrap current (see [17]). This current, which is present in all tokamak plasmas that have density and temperature gradients, is vital for the economic steady state operation of a tokamak, as it reduces the current that has to be driven externally by the solenoid or other sources. However, as we will see in the next section, it has the disadvantage that the current gradients that are produced drive plasma instabilities.

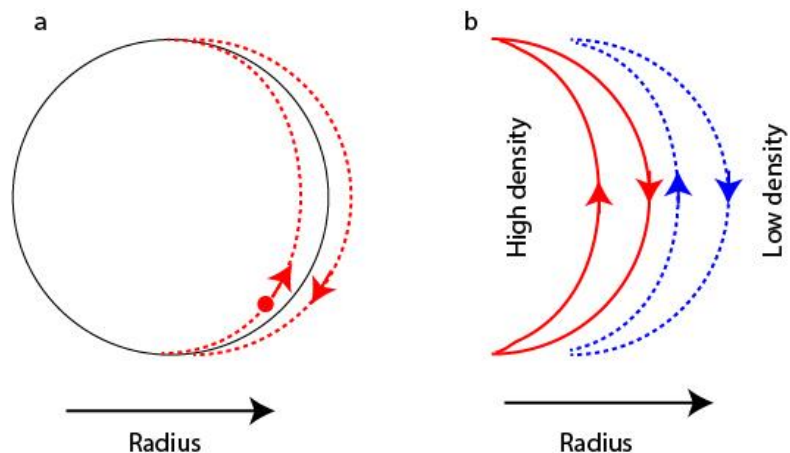

**Figure 4** Illustrations of a) Banana orbits of so called "trapped particles in a tokamak and b) the effect of a density gradient.

## 2.4 How to make a tokamak fusion plasma and different confinement modes

The previous section described the magnetic equilibrium of a tokamak plasma, in this section we look at how the plasma is actually formed. First of all a quantity of neutral gas, normally deuterium, is injected into the vacuum vessel. An electric field is induced by rapidly increasing the current in the solenoid. This electric field is strong enough to ionise the gas leading, typically in less than 10ms, to the formation of a relatively cold low current plasma which is captured in a pre-arranged magnetic field configuration. By ramping the current in the solenoid the plasma current is increased and by ohmic heating the plasma is heated, initially to temperature of a few million degrees. The vertical magnetic fields are adjusted to shape the plasma and the divertor current is applied so as to pull the plasma away from vessel components such that it only interacts with the divertor structures. At this point additional heating sources are applied to the plasma. These heating systems include neutral beam heating [18] or radio frequency heating that is often based around the electron or ion cyclotron resonance frequency [19]. While it is possible to heat the core of such a plasma to high temperatures, turbulent structures at the edge of the plasma lead to a loss of particles and these limit the confinement. An example of an image during a period when the plasma is in a so-called "Low Confinement" mode (L-mode) is shown in Figure 5a. The image was obtained in the Mega Ampere Spherical Tokamak (MAST) [20] located at the Culham Centre for Fusion Energy, UK. In a spherical tokamak the plasma shape resembles a cored apple rather than the traditional doughnut-shape in more traditional devices. Whilst the tokamak physics is the same in all these devices the spherical geometry affords excellent diagnostic visibility, which makes it an ideal device to demonstrate the phenomena discussed in this paper.

The light observed in Figure 5a is mainly due to $D_\alpha$ line radiation which is produced by the interaction of electrons with temperatures in the range of 10-100 eV with the neutral deuterium atoms that surround the confined plasma. The blurry edge of the plasma is due to turbulent eddies that are producing a large amount of transport at the plasma and reducing the confinement. This is analogous to steam carrying off energy from the open saucepan of boiling water that we discussed in section 1.1.

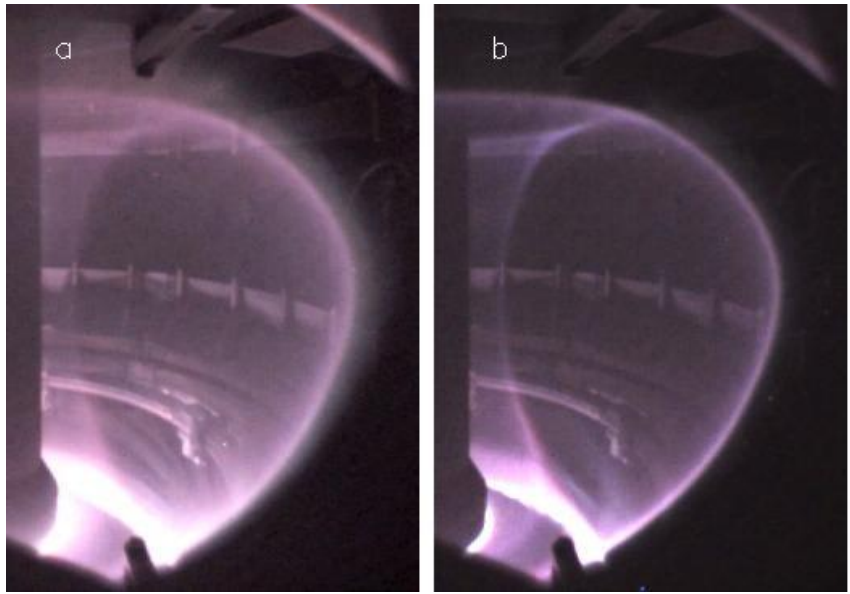

**Figure 5** Visible images of a tokamak plasma in MAST in a) L-mode and b) H-mode confinement regimes,

However, when the input power is raised above a critical level the plasma spontaneously organises itself into an improved or "High confinement" (H-mode) regime (see [21] and references therein). In this regime the edge turbulence is drastically suppressed, as evidenced by the sharp edges of the plasma that are visible in Figure 5b. A strong flow shear at the plasma edge develops in H-mode, which leads to the breaking up of the turbulent eddies [21]; a similar effect is observed in Jupiter, where the band structures are sustained due large velocity gradients between the bands (see for example [22]). The suppression of the turbulence leads to a decrease in particle and energy transport and an edge transport barrier is generated. The transport barrier acts in a similar way as the lid on the saucepan. The barrier leads to a steep pressure gradient at the plasma edge. Figure 6 shows the radial pressure profiles during the L and H-mode phases of this MAST plasma. In the H-mode phase the edge transport barrier leads to a steep gradient being formed at the plasma edge (in the region of normalised radius between 0.95 and 1.0). This increase in edge pressure also produces an increase in the core pressure; it is as though the plasma pressure has been raised – or sat on a "pedestal", which gives this edge region its name. This increase in overall pressure leads to an increase in overall confinement of a factor of two.

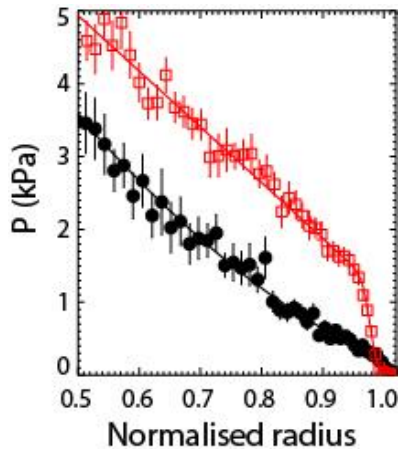

**Figure 6** The plasma pressure as a function of normalised radius for an L-mode (solid circles) and H-mode plasma (Open squares) from a MAST plasma.

However, this gift from the plasma of enhanced confinement does not come for free. As was described in section 2.3, this pressure gradient in the pedestal region, through the bootstrap current mechanism, results in currents and current gradients near to the edge of the plasma. These combine to lead to explosive plasma edge instabilities called Edge Localised modes (ELMs), which will be discussed in the next section.

### 3. Explosive plasma edge instabilities - Edge Localised Modes (ELMs)

Time traces of the $D_\alpha$ light, the stored energy and the electron density of a MAST plasma that undergoes a transition from L to H-mode are shown in Figure 7. At 0.205 s there is a sudden drop of the $D_\alpha$ radiation, signifying the transition from L-mode to H-mode; this is due to the decrease in the number of particles leaving the edge of the plasma and interacting with the surrounding neutrals. At the same time the stored energy in the plasma starts to increase, as does the density, due to the improved confinement. At 0.215 s, and intermittently thereafter, there is a burst of $D_\alpha$ light which causes a drop of ~ 4 % in the density. Note; the stored energy shown does not reflect this drop as it is calculated on a slower timescale − if it were available on a similar timescale it would have shown a drop of ~ 5%. Each peak in the $D_\alpha$ intensity is due to an ELM, which is a repetitive instability associated with the steep pressure and current gradients, which have formed at the plasma edge [23][24].

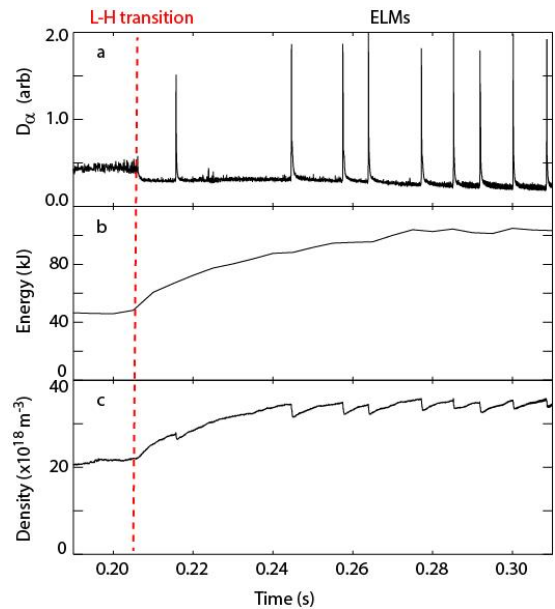

**Figure 7** Time traces of a) the $D_\alpha$ light emitted by the plasma, b) the energy stored in the plasma and c) the plasma density for a discharge on MAST that transitions from L to H-mode.

There are two basic ideal MHD instabilities associated with ELMs: the ballooning mode and the peeling mode (see [25] and reference therein). The peeling mode is associated with the edge current density, while the ballooning mode is driven by the pressure gradient. While each type of instability can lead to a different type of ELM, the largest one, the so called "type I" ELM, is caused by the overlap of the two instabilities – a peeling-ballooning mode. Like solar eruptions, type I ELMs are explosive events, which eject large amounts of energy and particles from the confined region [24]. Type I ELMs result in the sudden release of 5-15 % of the energy stored in the plasma in a short amount of time (100-300 µs), which results in large heat fluxes to plasma facing components [26].

The peeling-ballooning mode theory predicts that the edge current density and pressure gradient grow in the period between the ELMs until the peeling-ballooning stability boundary is crossed at which point the ELM is triggered (see Figure 8). The stable and unstable regions can be calculated with ideal MHD stability codes. An example of such a code is ELITE (Edge Localized Instabilities In Tokamaks Equilibria), which performs intermediate-to-high toroidal mode number (n) MHD stability analysis of the tokamak edge transport barrier region [27]. The location of the stability boundary is found to be in good agreement with the experimental parameters measured at the onset of an ELM. An example of the stability boundary calculated from this code is shown in Figure 8 together with the experimentally determined edge pressure gradient and current density as a function of time between two ELMs. At the beginning of the ELM cycle both the pressure gradient and current density are low and the edge parameters sit in the stable region. The pressure gradient and edge current steadily increase with time until they reach the peeling-ballooning stability limit at which point the ELM is triggered, which reduces the edge pressure and current and the cycle starts again.

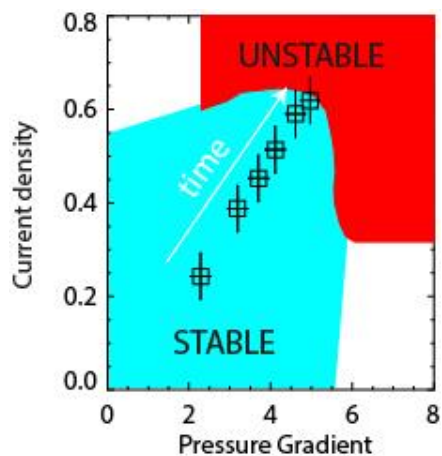

**Figure 8**  Current density and pressure gradient at the plasma edge as a function of time in an ELM cycle showing stable and unstable regions.

The peeling-ballooning model predicts that type-I ELMs in tokamaks will onset due to an instability that has a certain toroidal periodicity. This periodicity is classified in terms of a so called "toroidal mode number", which is typically in the range 10-20 [28]. The Wilson and Cowley model [29] of the ELM predicts the explosive growth of this mode. The structure of the mode is elongated along a field line, localised in the flux surface, perpendicular to the field line and relatively extended radially. The mode grows explosively as the time approaches a "detonation" time when the theory predicts the explosive growth radially of narrow filaments of plasma, which push out from the core plasma into the Scrape Off Layer (SOL). Such filament structures have been observed experimentally; initially using visible imaging on MAST [30] and subsequently using a variety of diagnostics on a range of devices (see [31] and references therein). These filaments subsequently separate from the edge of the plasma and travel out radially towards the vacuum vessel wall, carrying with them particles and energy.

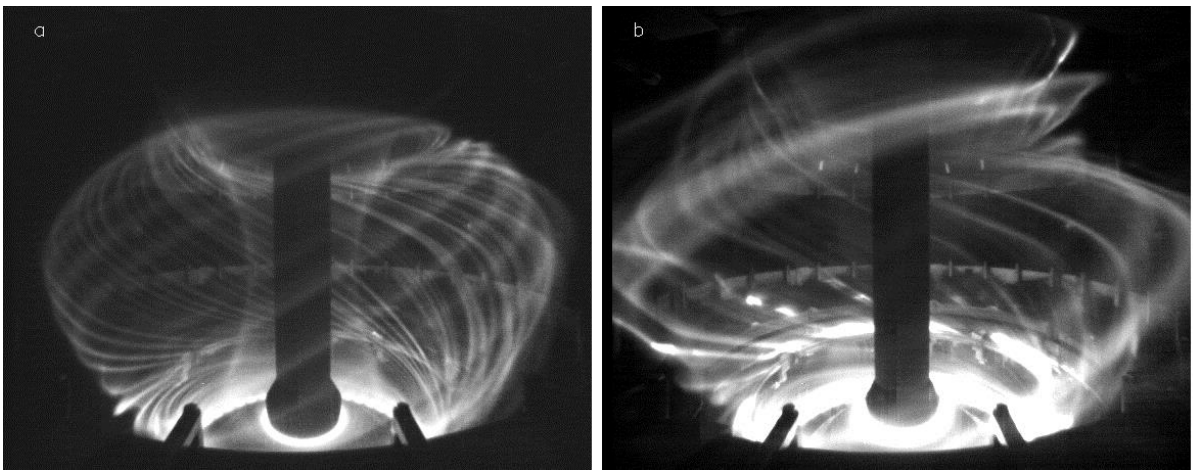

**Figure 9** Visible images captured on MAST using a 5 μs exposure time a) at the start of and ELM and b) during the eruption of the filament like structures.

An example of the observation of these filaments once they have pushed out into the SOL is shown in Figure 9a, which shows a wide angle view of a MAST plasma obtained using high speed visible imaging just after the start of the rise of the target $D_\alpha$ light associated with an ELM. Clear stripes are observed, which are on the outboard or low field side (LFS) edge of the plasma. The toroidal mode number can be calculated by counting the number of discrete filaments, which in this case is ~15. About 50-100 μs later the filaments separate from the plasma edge and start to travel radially (Figure 9b).

Several non-linear codes have been developed in recent years to study the crash phase of ELMs. Considerable progress has been made and the codes are approaching realistic plasma parameters and are starting to see experimental phenomena and trends (see [32] and references therein). These non-linear simulations have now reached a stage where they can be compared in detail with experimental data (see for example [33][34]). The simulated filament size and energy content are similar to the experimental observed ones [34] but the filaments are often more regularly spaced than in the experiment. This is possibly due to the fact the simulations are often performed with only a single mode number (see Figure 10), whereas a real ELM may be due to the interaction of several modes.

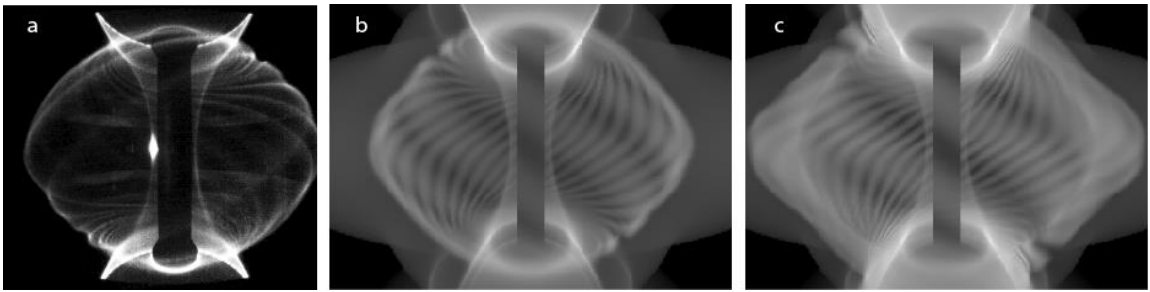

**Figure 10** Comparison of a) an image of an ELM on MAST obtained within 20 μs of the onset of the ELM with images produced by the JOREK code at b) 20μs and c) 80 μs after the ELM onset.

Whilst considerable agreement has been achieved, the ability to accurately predict ELM sizes from first principles has not yet been demonstrated. In all these cases the non-linear simulations of ELMs start from an MHD unstable state i.e. the pedestal profiles have to be increased beyond the peeling-ballooning stability limit. The ELM size depends on the initial linear growth rate and hence how far above stability the simulation is started and this makes it difficult for them to reliably predict the ELM amplitude. Therefore, scaling from present day experiments is used to estimate the ELM size on future devices.

Although the energy arriving at the divertor can be tolerated on today's devices, extrapolations of the ELM energy size to future larger scale devices show that this will not be the case. For example, the next generation machine, called ITER (originally an acronym of International Thermonuclear Experimental Reactor) [35], which is currently under construction in France, aims to produce a fusion gain factor $Q\sim10$ (where $Q$ is ratio of the power produced by the fusion reactions divided by the input power) in a plasma scenario that has a plasma current ($I_P$) of 15 MA. In such a scenario the expected natural ELM frequency is $\sim$ 1Hz with each ELM releasing ($\Delta W_{ELM}$) $\sim$ 20 MJ from the core plasma. This energy will be deposited on the divertor target in $\sim$ 500 μs giving large peak power fluxes that may be sufficient to melt the target material [36].

In order to ensure an adequate lifetime of the divertor targets the maximum ELM energy flux that can be repetitively deposited is 0.5 $MJm^{-2}$ [37]. Combined with assumptions on the radial energy deposition profiles this sets a maximum energy lost from the plasma during an ELM of $\Delta W_{ELM} = 0.66$ MJ [38].

Hence some form of either removing the ELMs or making them smaller, so called ELM mitigation, will be required. In the next section we will discuss methods that have been shown to be able to do this and hence to tame the ELM.

## 4. ELM taming

All current estimations of the energy released by type I ELMs indicate that, in order to ensure an adequate lifetime of the divertor targets on ITER, a mechanism is required to decrease the amount of energy released by an ELM (ELM mitigation), or to eliminate ELMs altogether (ELM suppression) [39]. It is essential that whatever technique is used to do this, must also retain most, if not all, of the improved confinement associated with the H-mode. To work out just how much mitigation is required we need to know how the ELM size scales with frequency and how this scales with plasma parameters.

Firstly, it has been observed on all devices that the ELM size ($\Delta W_{ELM}$) multiplied by the ELM frequency ($f_{ELM}$) remains a constant fraction of the input power ($P_{in}$) i.e. $\Delta W_{ELM} x f_{ELM} = 0.3\text{-}0.4 x P_{in}$ [40]. Hence increasing the ELM frequency will help. ITER will operate at a range of plasma currents ($I_P$), with the highest fusion power expected in the highest $I_P$ discharges. The ELM frequency scales with plasma current as $f_{ELM} \propto I_P^{-1.8}$ [40], so the natural ELM frequency in ITER will vary from $\sim$ 1 Hz for plasmas with $I_P =$ 15MA to $f_{ELM}$ ~7Hz for $I_P$=5 MA. So a higher level of mitigation will be required for the higher plasma current discharges. Combining all this information and taking into account the changes in the power deposition profile and the sharing between targets, a mitigated ELM frequency required to keep the divertor energy flux density below the 0.5 MJm$^{-2}$ limit can be calculated as a function of $I_P$ [41]. These calculations show that for discharges with $I_P$>8MA some form of ELM mitigation (increase in ELM frequency over the natural value) is required.

Although ELMs have a deleterious effect on the divertor, it has been found that ELMs can play a beneficial role since they help flush out impurities i.e. any ions that are not the D or T fuel. The interaction of the edge plasma with the surrounding surfaces leads to the generation of impurities through sputtering (either physical or chemical) [42]. In ITER the divertor will be made out of tungsten (W) to manage the power and particle loads. If the W released from the divertor got into the plasma and accumulated there then it would significantly reduce the fusion yield, since a concentration of heavy impurities can radiate significant energy from the plasma [43], causing it to cool. In order to avoid any problems it is necessary to ensure that the W concentration remains below $2.5 \times 10^{-5}$ of the electron density [41]. It has been observed on current devices that ELMs, provided their frequency is high enough, are very effective at expelling high Z impurities from the edge of the plasma, which avoids accumulation in the core (see [41] and references therein). This leads to a requirement that the minimum ELM frequency on ITER, irrespective of plasma current, is ~ 18Hz.

Combining the requirements of avoiding damage to the plasma facing components and W accumulation in the core results in the required increase in ELM frequency over the natural ELM frequency, as a function of $I_P$, to be in the range ~3-40 [44]. Although ELMs associated with such an ELM frequency are predicted to be below the damage threshold for the divertor target, they would still produce a large thermal cycling of the target materials. The complete removal of the ELMs, called "ELM suppression", may in fact be the best solution for the divertor but the mechanism that results in the suppression must also be associated with sufficient impurity transport to avoid tungsten accumulation.

Several ELM control techniques have been investigated for ITER (see [45] and references therein). These include:

1) Firing frozen deuterium pellets into the plasma edge; each pellet triggers an ELM and hence can be used to pace the ELM frequency.

2) Applying a vertical kick to the plasma; provided the vertical displacement is large enough an ELM is triggered.

3) The application of magnetic perturbations; we want to keep the good confinement due to H-mode but need to stop the instability associated with the ELM growing too large and so need to produce either rapid small ELMs or no ELMs at all. The aim is to modify the plasma near to the plasma edge, while keeping the core confinement unchanged. This is done by modifying the flux surfaces at the plasma edge using a non-axisymmetric perturbation to the magnetic field.

This last technique is the subject of the remainder of this paper.

## 4.1 The application of non-axisymmetric magnetic fields

The underlying idea for using magnetic perturbations to suppress ELMs is based around the fact that the Edge Transport Barrier (ETB) of a H-mode plasma is too efficient, i.e. the pressure gradient and the associated bootstrap current density can become too large. If, by some process, transport in the ETB could be enhanced just enough then the plasma would remain in the stable region with respect to Peeling-Ballooning instability (shown in Figure 8) and ELMs could be avoided.

In order to produce the nested flux surfaces described earlier and optimize the confinement, the magnetic fields need to be as toroidally uniform as possible. Tokamaks are traditionally built such that the magnetic field, at a particular r-z location, varies by less the $10^{-4}$ as a function of toroidal angle i.e. typically by less than 1 G. This puts stringent limits on the design, build and installation of the coils. For example, a tilt of ~ 0.1° in one of the coils used to produce the vertical fields can produce a so called intrinsic error field that can severely limit the operation of a tokamak (see [46] and references therein). In ELM control by magnetic perturbations the toroidal symmetry is broken on purpose using toroidally discrete coils. As an example, the coils used in MAST are shown in Figure 11. The coils are typically arranged in two rows above and below the mid-plane of the plasma and are powered individually such that the direction and magnitude of the current in each coil can be modified so as to optimize their effect. These coils produce a radial perturbation in front of them. The application of such perturbations effectively bend the field lines that pass in front of them in the radial direction, which can enhance the radial transport. It is known in particular that radial Resonant Magnetic Perturbations (RMPs), where "resonant" means that the perturbation is aligned along the pre-existing equilibrium magnetic field line, can "tear" the nested flux surfaces, creating so-called "magnetic islands" [8]. When a magnetic island is present inside the plasma, the radial transport is vastly increased, as the island short circuits the flux surfaces.

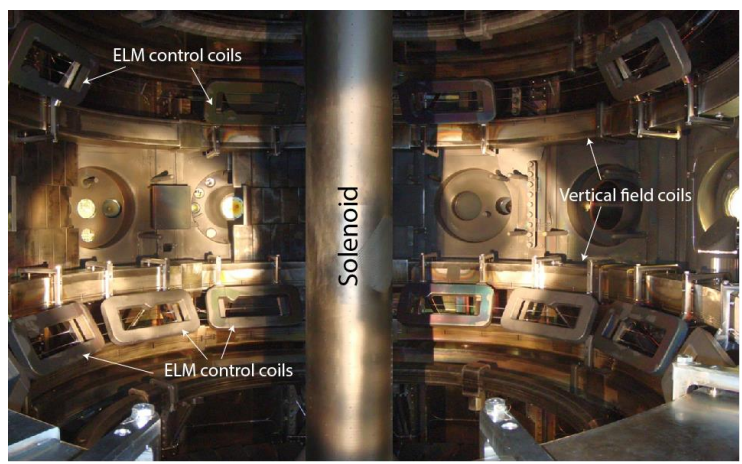

**Figure 11** Photograph of the inside of the MAST tokamak showing the location of the ELM control coils.

To see how the magnetic perturbations work let us return to the picture of flux surfaces shown in Figure 3 and described in section 2.2. Imagine breaking up a single flux surface into say 1000 discretised points. Then choose a given toroidal position and follow that point around the tokamak and plot the location in the poloidal plane after 1 turn. Repeating this for each of the 1000 points would then produce a so called "Poincaré plot". Figure 12a shows what happens if this is done for the axisymmetric case, without the application of magnetic perturbations. The resulting image is effectively the flux surfaces shown in Figure 3a. This is then repeated with a perturbation in the localized RMP coils with a value of $10^{-4}$ of the toroidal field. In this case field lines that do not pass near the coil will be unaffected but field lines that pass straight in front of the coil will get a large radial displacement. The resulting Poincaré plot is shown in Figure 12b. The magnetic field lines start to behave in a chaotic manner, the flux surfaces near to the edge of the plasma are now destroyed and small magnetic islands are formed, however, the flux surfaces in the core remain unaffected. If the perturbation strength is doubled, then the radial depth to which the flux surfaces are destroyed increases (Figure 12c). The edge region is now often referred to as "stochastic" or "ergodic" (see [47] and references therein for a fuller description).

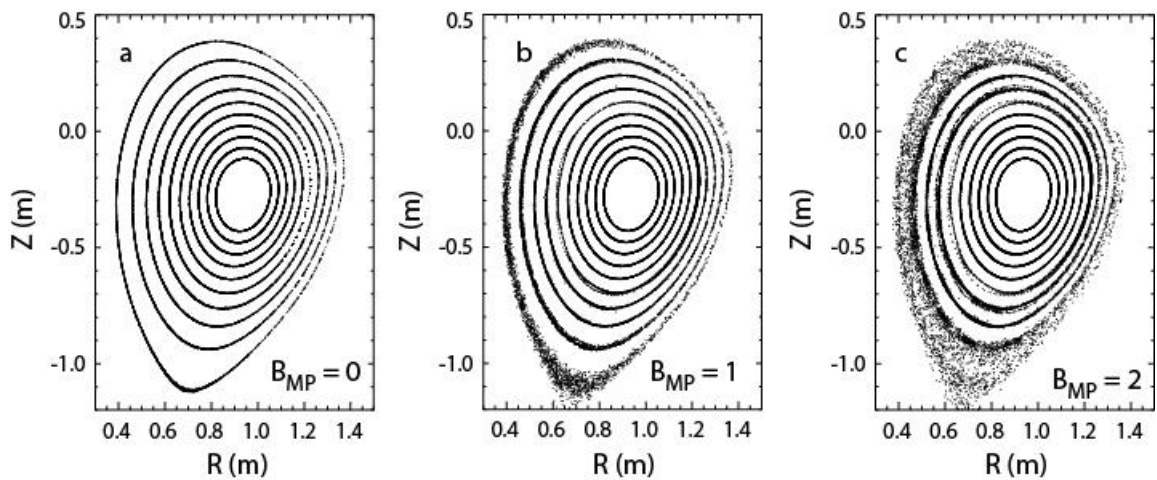

**Figure 12** Field line tracing to produce Poincaré plots for a) no RMP field and with an RMP with a field strength of b) 1 and c) 2 in units of $10^{-4}$ of the toroidal field.

As discussed in section 2.3, in an axi-symmetric diverted tokamak the Last Closed Flux Surface (LCFS) separates the region of confined and open field lines. If a field line on the LCFS is followed around the tokamak in one direction and then in the other the two surfaces generated overlap. However, if the same is repeated for a case when the non-axi-symmetric magnetic perturbations are applied the surfaces do not overlap and instead form a pair of so called "manifolds" [48]. Structures are formed where the manifolds intersect and these are particularly complex near to the X-point. The manifolds form lobes that are stretched radially both outwards and inwards.

Calculations of what these lobes should look like can be performed by again carrying out field line tracing using the 3D magnetic field, which is the sum of the external magnetic perturbations from the RMP coils and the equilibrium plasma fields. Such a combination of the fields is referred to as vacuum modelling since it assumes that the plasma equilibrium does not respond to the applied perturbation. Again we start off by choosing discrete field lines, this time in the X-point region. These field lines are traced in both directions until they either leave the plasma or complete 200 toroidal turns. For each field line traced the furthest depth that it reaches inside the plasma is recorded [49]. A plot is then made in the poloidal plane of the original location of the field line and the depth expressed in terms of normalised flux ($\Psi_N$), where $\Psi_N = 0$ is defined as the centre of the plasma and $\Psi_N = 1$ is the LCFS.

Figure 13a shows the resulting plot at a given toroidal angle when no RMP is applied; as expected regular contours are observed. These are   consistent with the expected flux surfaces. Figure 13b shows the effect of the applying the RMPs with a 6-fold periodicity in the toroidal direction; the perturbation is said to have an n=6 toroidal mode number. The resulting figure shows clear lobe structures that protrude from the plasma edge and contain field lines which originate from inside the plasma. Such lobe structures have been observed experimentally and, as we will discuss below, reveal a lot about how the applied perturbations interact with the plasma.

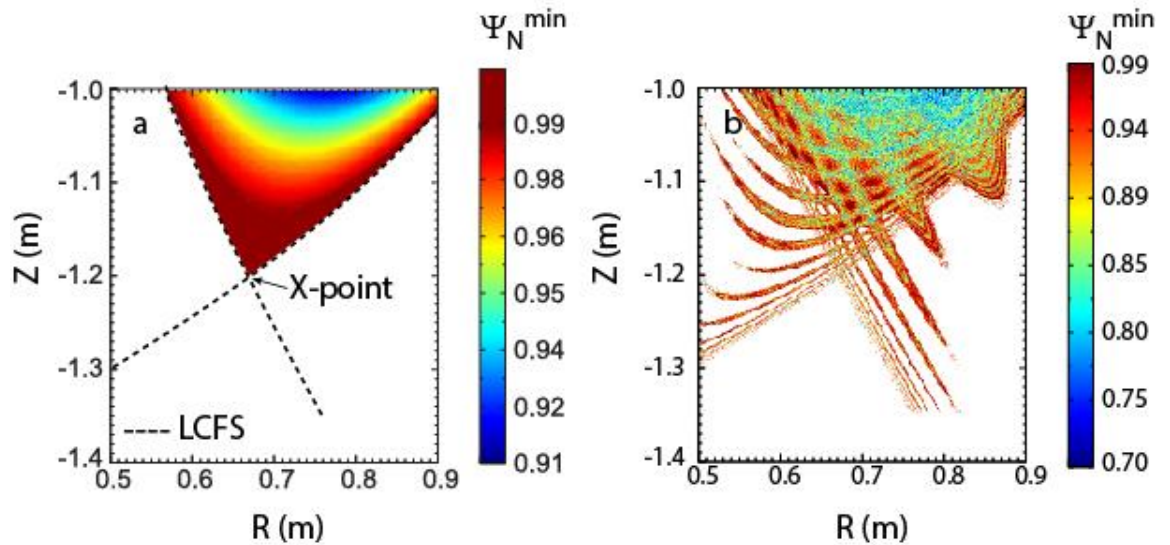

**Figure 13** Poincaré plots from vacuum modelling showing the minimum value of $\Psi_N$ reached by a field line a) without RMPs and b) with RMPs in a n=6 configuration.

## 4.2 Lobe structures

The lower X-point region of the MAST plasma has been imaged using a toroidally viewing camera with a spatial resolution of 1.8mm. The image has been filtered with a $He^{1+}$ (468 nm) filter using an integration time of 2 ms (for more information on the technique used see [50]). This spectral line has been chosen since it is localised near to the LCFS region for the typical plasma conditions found in MAST. Figure 14a shows what is observed when no RMPs are applied; A smooth boundary layer associated with the LCFS is observed. In contrast, Figure 14b shows an image obtained when the RMPs are applied; Clear lobe structures are seen near to the X-point.

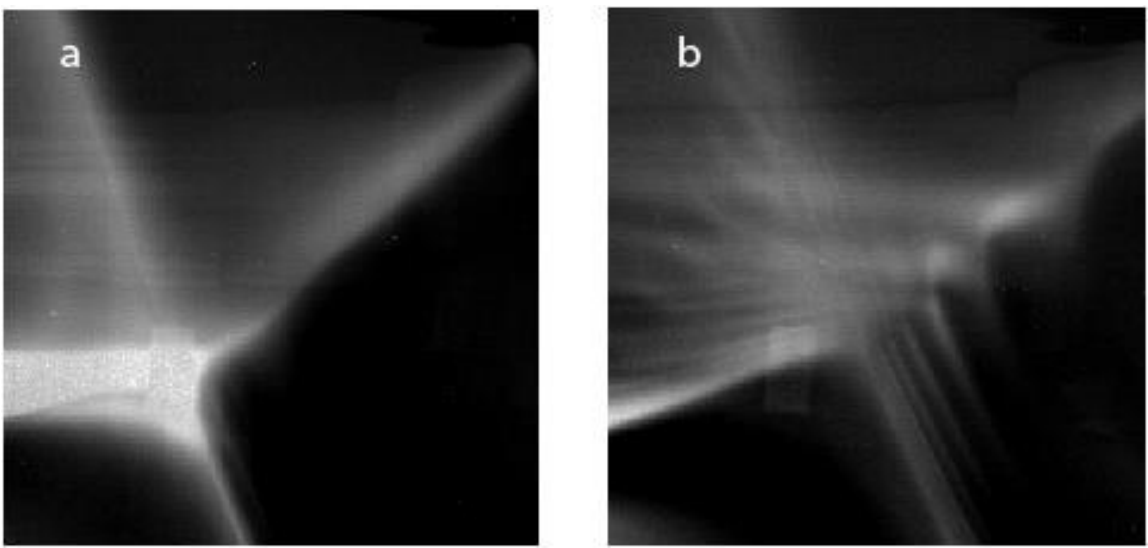

**Figure 14** $He^{1+}$ filtered images captured between ELMs a H-mode plasma a) without RMPs and b) with RMPs in an n=6 configuration.

When the modelled data shown in Figure 13b was mapped onto the image shown in Figure 14b taking into account the viewing location, a good quantitative agreement was obtained between the number and separation of the lobes, however, there appeared to be a discrepancy in their radial extent [51], with the lobes observed in the experiment being shorter. This discrepancy could be due to: 1) the possibility that we are not seeing all the lobes due to sensitivity to the distribution of the $He^{1+}$ emission or 2) the fact that the applied fields cannot simply be combined with the equilibrium field.

To investigate the effects of the $He^{1+}$ emission a forward model of the camera data was constructed [50]. Data recorded from a shot without RMPs were used to generate a map of $He^{1+}$ light emission as a function of the flux surface quantity $\Psi_N$. The results from the field line tracing shown in Figure 13b were then used to determine the light emission within the lobes. The resulting simulated image is shown in Figure 15a. The radial extent of the simulated lobes is again too large and hence we need to understand how the applied fields actually interact with the plasma.

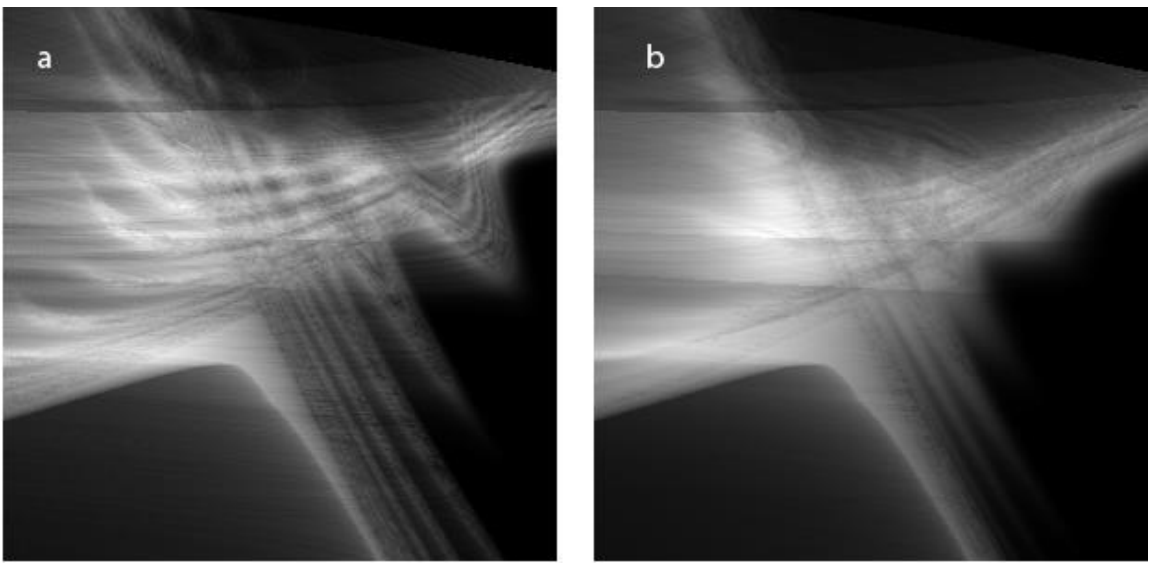

**Figure 15** Simulation of the $He^{1+}$ light emission for a MAST discharge using a) unscreened and b) screened RMPs in a n=6 configuration.

### 4.3 Screening of applied fields

To date in the discussion on the application on RMPs, we have assumed that we can just add the perturbation field to the pre-existing plasma field. However, in the discussion of calculating the equilibrium in section 2.2 we introduced ideal MHD in which the plasma is considered to be a perfect conductor. In a perfect conductor, the interior magnetic field must remain fixed and if a field is applied the conductor will generate currents to prevent any change in magnetic flux passing through it. A perfectly conducting plasma does this by setting up electric currents such that the applied field is zero on field lines that join on themselves after $m$ toroidal turns and $n$ poloidal turns, i.e. where q=m/n has integer values of *m* and *n*. The magnetic fields produced by these surface currents cancel the applied magnetic field exactly at these surfaces. The difference between a perfect conductor and a superconductor is that in a superconductor, the magnetic field is always zero within the bulk of the superconductor whereas in a perfect conductor, whilst the magnetic field must remain fixed, it can have a non-zero value.

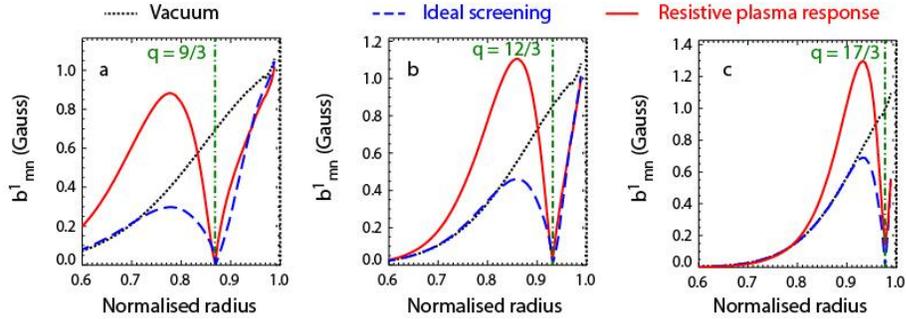

**Figure 16** Radial profile of the applied perturbation field Fourier decomposed in terms of the poloidal (m) and toroidal mode numbers for an RMP in the vacuum approximation (dotted), from an ideal screening code (dashed), from resistive MHD (solid) with n=3 a) m=9, b) m=12 and c) m=17. The vertical dashed-dotted line shows the radial location of the relevant rational surface.

Since the screening currents are localized to the layers within the plasma where q=m/n is rational, we first need to Fourier decompose the applied magnetic perturbation field in terms of *m* and *n* to produce $b^1_{mn}$ [52], where n is the toroidal mode number of the applied perturbation. Figure 16 shows the radial profiles for several Fourier components resulting from the application of a magnetic perturbation with a toroidal mode number n=3. The dotted line in Figure 16a shows the component corresponding to m=9 n=3 ($b^1_{93}$) in what is called the vacuum approximation i.e. where it is assumed that the magnetic perturbation can just be added to the plasma equilibrium field. This total field component falls off smoothly with distance into the plasma. The dashed-dotted vertical line shows the location in the plasma of the q=9/3 surface which is located at a normalised radius of 0.87. Ideal MHD would say that the resonant field at this location must go to zero i.e. $b^1_{93}$=0 at the q=9/3 surface and this is what can be seen from the dashed line in Figure 16a which has been calculated using a screening current code [53]. The plasma screens out this component of the applied field by producing a screening current that flows on this rational surface. As well as cancelling the applied field exactly at the rational surface it also modifies the field over a wider radial extent. Figure 16b and c show other examples for the m=12 and m=17 components also, where again perfect screening is achieved at the respective rational surface. If the plasma were a perfect conductor then no resonant field would penetrate and no lobes would be observed. The observation of the lobe structures proves that some field penetration occurs.

The screening currents arise due to the flow of the electrons in front of the applied perturbation field. Hence the screening currents can only flow where the resistivity of the plasma is sufficient low and where the electrons have a non-zero velocity. Near the plasma edge the temperature is lower, typically $\leq 1$ keV, and here the non-zero resistivity means that the screening currents that can be produced do not exactly cancel out the applied field. In addition, in the steep pressure gradient region the currents required to balance this gradient produce what is called a "diamagnetic flow" of the electrons. This flow can exactly cancel the bulk flow of the plasma and hence in the rest frame of the applied perturbation it is possible that at a particular location the electrons are at rest (see [54] and references therein). This typically only occurs over a very narrow radial region of the plasma but if it exists it is located at or near the pedestal top. At such a location, screening currents cannot be generated and hence the applied perturbation field will exist at this location irrespective of the plasma resistivity.

In ideal MHD the plasma is tied to the magnetic field lines and although radial fields can bend these field lines they cannot be broken and hence the magnetic islands that we discussed earlier could not be created. Basically in ideal MHD the field lines would just get tied up into knots. An extension of ideal MHD is resistive MHD, which can be used to take into account all these features self-consistently. Several computers codes have been developed to perform these calculations. The solid curves in Figure 16 show the results of such a calculation using the MARS-F code [52], which is a linear single fluid resistive MHD code that calculates how the plasma responds to the applied perturbations, including screening effects due to toroidal rotation.

In order to simplify the understanding we have chosen the plasma parameters such that there is no region in which the electron velocity is zero. The $b^1_{mn}$ component still goes to zero at the q=9/3 and q=12/3 surfaces as the plasma resistivity is sufficiently low here. Nearer towards the plasma edge the resistivity increases and full screening is not achieved and hence the perturbation field does not go to zero at the q=17/3 surface. In contrast to the simple screening current model, away from the rational surfaces the resistive MHD calculations show an amplification of the applied field. This is due to the fact that the application of the applied field causes what is called a "plasma response" – the plasma readjusts itself in a 3D manner – displacing field lines to produce a minimum energy solution that has a modified magnetic field structure. These more complete calculations can now be included into the lobe simulations resulting in the image shown in Figure 15b. The radial extent and width of the lobes is now in good agreement with the experimental image shown in Figure 14b.

The screening currents are not aligned with the equilibrium magnetic field and hence create a $\vec{j} \times \vec{B}$ torque on the plasma. This torque acts to brake the toroidal rotation of the plasma [55]. In addition to the good agreement found in comparing the lobe sizes the calculated torque on the plasma is also in good agreement with the rotation braking observed [56] indicating that there is a good understanding of how the fields penetrate and are screened.

### 4.4 Effect of RMPs on ELMs

The use of Resonant Magnetic Perturbations (RMPs) has been employed to either make ELMs more frequent and smaller (ELM mitigation) or to suppress type I ELMs on a range of devices; DIII-D [57][58], JET [59], MAST [60], ASDEX Upgrade [61] and KSTAR [62]. A lot of the early research was performed on the DIII-D device, which demonstrated suppression of type I ELMs in a plasma with a similar shape and normalised edge parameters to the Q=10 ITER baseline scenario [58]. The subsequent worldwide activity and confirmation on different devices has led to a set of in-vessel RMP coils being considered as one of the two main systems for ELM control in ITER [63].

Figure 17 shows an example of the effect that the application of RMPs with toroidal mode number n = 2 have on a H-mode plasma on the ASDEX Upgrade tokamak which is based at the Max Planck institute in Garching, Germany. After the coil current (Figure 17a) reaches a certain threshold the large type I ELMs, which initially have a frequency of 60 Hz, first become more frequent and smaller (i.e. are mitigated) and then disappear altogether and are replaced with very high frequency fluctuations. There is also a drop in the density (Figure 17b) due to the enhanced transport associated with the RMPs, which moves the plasma edge parameters to a region in which they are stable to type I ELMs. There is also a reduction in the toroidal rotation velocity (Figure 17c) due to the generation of the screening currents discussed previously.

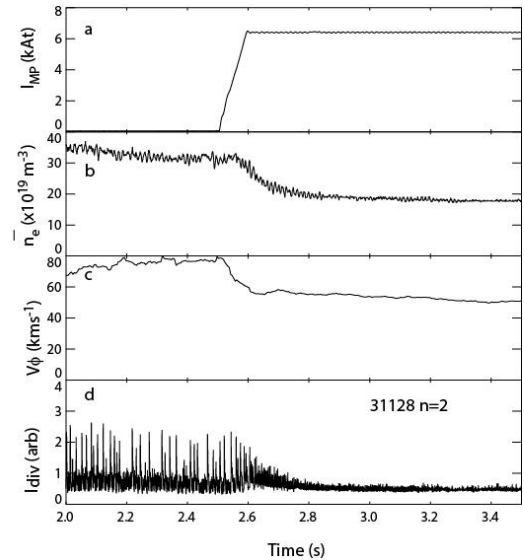

**Figure 17** Results from ASDEX Upgrade: a) the current in the magnetic perturbation coils ($I_{MP}$) b) plasma density, c) the toroidal rotation at the top of the pedestal ($V_\phi$) and d) the divertor intensity for a discharge with RMPs in a n=2.

So the technique clearly works – it reduces both the ELM energy losses and the heat loads to the divertor targets [64]. However, there is one drawback and this is associated with the enhanced particle transport and the reduction in the plasma density, which leads to a drop of overall confinement. While the confinements is still much better than in L-mode, any reduction will mean that electricity from future fusion power stations will be more expensive, so there is an ongoing active area of research aimed at reducing this loss. There have been some recent successes [64] but more work is required.

## 5. Summary and conclusions

In this paper we have discussed the potential that nuclear fusion reactions, using deuterium and tritium as fuel, have for solving the Earth's energy needs. All DT fusion techniques currently being studied as an energy source require the fuel to be heated to temperatures in excess of 100 million degrees, which means that the fuel ends up in the form of a plasma. While stars use gravity to confine this plasma on Earth other techniques are required. The techniques that are currently being actively researched are inertial and magnetic confinement.

The various aspects of the magnetic fields required to confine a plasma have been presented and, in particular, a detailed description of the tokamak, which currently has achieved the largest confinement times has been discussed. In a tokamak the plasma reaches an equilibrium state, where the pressure gradients are balanced by currents flowing in the plasma. However, this equilibrium is not steady state and is continually evolving due losses of heat and particles from the plasma being balanced by external heating and fuelling. As the heating is increased, the plasma spontaneously re-organises itself into a higher confinement regime. Whilst this regime is very good for confinement, the edge transport barrier that leads to the improvement is, if anything, too efficient and leads to sharp pressure and current gradients at the plasma edge.

These gradients at the edge of the plasma lead to explosive instabilities called ELMs. While ELMs are good at removing impurities from the plasma edge, if the energy loss associated with the ELM is too large it will reduce the lifetime of the plasma facing components in future reactors. Current predictions show that in the next step tokamak, ITER, which is currently under construction in France some way of controlling or taming ELMs will be required. Several techniques are currently being investigated and one of the leading contenders is the application of toroidally asymmetric magnetic perturbations.

To understand how magnetic perturbations can control ELMs we first need to understand how these perturbations enter the plasma, which ideally is described as a perfect conductor. However, near the edge of the plasma, where the temperatures are low and hence the resistivity higher, this ideal picture breaks down. We have seen how measurements of the distortions to the plasma shape can be used to confirm our understanding of how the field penetrates and is screened.

Finally we have seen that the application of these magnetic perturbations can be used to either reduce the size of the ELM or in some cases remove them altogether. Although not discussed in detail in this paper, the 3D distortions to the plasma shape play a key role in determining what happens to the ELM frequency and energy loss due to the application of these perturbations [65]. The main aim of future research is to optimise the performance of these magnetic perturbations and to build sophisticated models that will allow extrapolation to future devices.

## Acknowledgement

This work has been (part-) funded by the RCUK Energy Programme [grant number EP/I501045]. To obtain further information on the data and models underlying this paper please contact PublicationsManager@ccfe.ac.uk.